\documentclass[a4paper,10pt]{article}
\usepackage[usenames,dvipsnames,svgnames,table]{xcolor}

\include{PhysNote} 
\usepackage{srcltx} 

\usepackage{amsmath,amsthm,verbatim,amssymb,amsfonts,amscd, graphicx}
\usepackage{graphics}
\usepackage{capt-of}
\usepackage{color}
\usepackage{url}
\usepackage{hyperref}
\hypersetup{colorlinks=true}
\usepackage{epsfig}
\usepackage{multirow}
\usepackage{booktabs} 
\usepackage{tikz}
\usepackage{pifont}
\usepackage{arydshln} 
\usepackage{mathtools}
\usepackage{parskip}

\def\be{\begin{equation}}
\def\ee{\end{equation}}
\def\ba{\begin{eqnarray}}
\def\ea{\end{eqnarray}}

\def\ga{\mathrel{\raise.3ex\hbox{$>$\kern-.75em\lower1ex\hbox{$\sim$}}}}
\def\la{\mathrel{\raise.3ex\hbox{$<$\kern-.75em\lower1ex\hbox{$\sim$}}}}

\newcommand{\fr}[2]{\frac{#1}{#2}}

\newcommand{\oo}{\omega_{0}}
\newcommand{\oa}{\omega_{a}}
\newcommand{\eff}{\rm{eff}}
\newcommand{\e}{\rm{E}}

\newcommand{\Omo}{\Omega_{\rm{m} 0}}

\newcommand{\Omn}{\Omega_{\rm{m}}(\overline{n})}
\newcommand{\Omz}{\Omega_{\rm{m}}(\overline{z})}

\newcommand{\m}{\rm{m}}

\newcommand{\PRD}{Phys.\ Rev.\ D}
\newcommand{\PRL}{Phys.\ Rev.\ Lett}
\newcommand{\PLB}{Phys.\ Lett.\ B}
\newcommand{\JCAP}{J.\ Cosmol.\ Astropart.\ Phys.}
\newcommand{\MNRAS}{Mon.\ Not.\ R.\ Astron.\ Soc.}

\newcommand{\DE}{\rm{DE}}

\newcommand{\wE}{\omega_{\rm{E}}}
\newcommand{\obs}{\rm{obs}}
\newcommand{\s}{\rm{s}}
\newcommand{\barn}{\overline{n}}
\newcommand{\barH}{\overline{H}}
\newcommand{\barz}{\overline{z}}
\newcommand{\bara}{\overline{a}}
\newcommand{\barg}{\overline{g}}
\newcommand{\bart}{\overline{t}}
\newcommand{\barR}{\overline{R}}
\newcommand{\barrho}{\overline{\rho}}
\newcommand{\HE}{H_{\rm{E}}}
\newcommand{\baromega}{\overline{\omega}}
\newcommand{\bargamma}{\overline{\gamma}}
\begin{document}

\baselineskip=16pt
\begin{titlepage}
\begin{center}

\vspace{0.5cm}

\large {\bf Conformal Frame dependence on Cosmological Observations in Scalar-Tensor Theories of Gravity}
\vspace*{5mm} \normalsize

{\bf Young-Hwan Hyun$^{\,1}$, Yoonbai Kim$^{\,1}$, Seokcheon Lee$^{\, 2}$} 

\smallskip
\medskip

$^1${\it Department of Physics, BK21 Physics Research Division, Institute of Basic Science \\ Sungkyunkwan University, Suwon 16419, South Korea}

$^2${\it The Research Institute of Natural Science, \\
Gyeongsang National University, Jinju 52828, Republic of Korea}

\smallskip
\end{center}

\vskip0.6in

\centerline{\large\bf Abstract}

Cosmological observations provide more accurate values both for background evolution of the Universe and for the structure formation. These values are given by the so-called dark energy equation of state, $\omega$ and the growth index parameter, $\gamma$. From these observed parameters, one can reconstruct the model functions in scalar tensor gravity theories. However, there is a long standing debate about the (in)equality between conformally transformed frames in scalar tensor gravity models. We show that cosmological observables are frame dependent when they are described by frame independent parameter, redshift. Thus, if the cosmological observables are interpreted in one frame, then all of the observables should also be interpreted in that frame. This explicitly shows the conformal inequality of cosmological observables. Also, our method provides the model independent analysis for STG models about various observables in both frames.

\vspace*{2mm}

\end{titlepage}

\section{Introduction}
\setcounter{equation}{0}
Despite the success of the theory of general relativity, the current observed accelerated expansion of the Universe invokes the existence of mysterious dark energy. An alternative explanation for the current acceleration of the cosmic expansion is a modification of general relativity to be a more general theory of gravity at a long-distance scale.  In essence, dark energy models modify the stress-energy tensor by adding an additional component with the present value of an equation of state (EOS) near a negative one, $\omega(z_0) \simeq -1$. The modified gravity theories correspond to modifying the four-dimensional Einstein–Hilbert action.

Even though the modified gravity theories modify the left-hand side (LHS) of the Einstein's field equations, the background evolutions of the spatially homogeneous and isotropic universe are described by Friedmann equations if one replaces the additional LHS term(s) with the effective dark energy \cite{08014606}. Thus, the expansion history of the Universe can be described by an effective dark energy EOS both for dark energy models and for modified gravity theories. For this purpose, Chevallier-Polarski-Linder (CPL) parametrization $\omega = \omega_0 + \omega_a (1 - a)$ is one of the suitable candidates to describe the general $\omega$ \cite{0009008,0208512}.

From this point of view, one is not able to distinguish models of different physical origins if they have the same background expansion history. Thus, one needs complementary methods beyond the background evolution of the Universe to break degeneracies among models. The growth rate of the large scale structure, $f(a) \equiv d \ln \delta_{m}(a)/ d \ln a$ is one of these discriminators of models \cite{0507184,0612452,07042421,07090307,08012431,08021068,08033292,10122646,13076619,160501644,170508797}.  This is due to the fact that the matter perturbations of the different gravity theories might have the different Poisson terms (source terms) for the same background evolution (friction term). 

It is often useful to parameterize the growth rate, $f$ by the so-called the growth index parameter (GIP), $\gamma = \ln f(a) / \ln \Omega_{\rm{m}}(a)$ both in theories and in observations. Thus, the different models have the different behaviors of $\gamma$ and it is also useful to adopt the various parametrizations of $\gamma$. Behavior of $\gamma$ in the so-called Dvali-Gabadadze-Porrati (DGP) braneworld model \cite{0005016,0010186,0105068} has been widely investigated \cite{08033292,0401515,0701317,08024122,09051735,09053444}. The GIP of generalization of Einstein-Hilbert action by a general function of the scalar curvature, named $f(R)$-gravity \cite{Buchdahl} has been also studied \cite{08032236,08093374,09035296,09082669}. In this model, $\gamma$ has also a scale dependence in addition to time dependence.  For the scalar-tensor theories of gravity (STG) \cite{Bergmann,Nordtvedt,Wagoner}, one can obtain the specific $\gamma$ for the certain STG model \cite{0001066,07101510,08024196}. One of the common parametrizations for all models is given by $\gamma = \gamma_0 + \gamma_a (1 - a)$ as similar to that of $\omega$ \cite{10122646}. 

STG is well motivated both for theory and for phenomenology. Theoretically, there exists a ubiquitous fundamental scalar field coupled to gravity in theories which unify gravity with other interactions \cite{Fradkin,Callan,Lovelace,Green,Polchinski}. Also the dynamical equivalence between the metric formalism f(R) theories and a particular class of STG has been shown \cite{Teyssandier,Magnano,9307034,0307338,09100434}. The same is investigated for the Palatini formalism f(R) gravity theories \cite{0308111,0604028}. Phenomenologically, STG naturally can explain several aspects. First, ``the lithium problem'' in the standard big bang nucleosynthesis (BBN) might be solved in STG due to its slower expansion than in general relativity before BBN, but faster during BBN \cite{0511693,0601299,08111845}. The weak lensing (WL) shear power spectrum in STG is different from that of GR due to the different growth history of the matter between two models \cite{0403654,0412120}. The Integrated Sachs-Wolfe (ISW) effect probes modified gravities on cosmological scales through the matter potential relation due to the effective anisotropic stress \cite{9906066,08032238,SLAIP,09092045,13081142}. Another benefit of STG is that the crossing phantom $\omega < -1$ preferred by observations can be naturally obtained \cite{0504582,0606287,0610092}. 

There have been a number of reconstructions of specific STG models which are consistent with known observational constraints \cite{0011115,0107386,0508542,0612569,07053586,08031106,10031686,10061246,10112915}. However, one needs to reconstruct theory without any specific theory {\it a priori}. Thus, we use both background and growth history parameters ($\omega, \gamma$) which can be obtained from observations to find the viable subclasses of STG \cite{10122646}. However, this process is done in the so-called Jordan frame (JF) where the metric tensor is minimally coupled to the matter sector. Based on this reconstruction, one also needs to investigate the background evolution in the Einstein frame (EF) by doing the conformal transformation of the metric. STG is a specific model because it can be interpreted as either a modified gravity theory in JF or an inhomogeneous dark energy model in EF. When one investigates the cosmological observables, e.g., the Hubble parameter and the growth rate in JF, one needs to convert the terms from the modified gravity into those of the effective dark energy. This effective dark energy is different from that in EF.

There has been a long-standing debate on physical equivalence of the two frames \cite{13081142,Magnano,Dick,9910176,0205187,0604492,0605109,10035394,160106152}. The conclusions are contradict to one another. The main discrepancies come from the fact that one uses the physical observables as a function of time which is the frame dependent quantity. Thus, one needs to describe the observables as a function of the redshift, $z$ which is the frame independent \cite{13081142}. If the cosmological observation values obtained from two different frames are different, then one needs to distinguish two frames. One still cannot declare which frame is physical, but one needs to use the one specific frame for analysis of observables.   

The background evolution and growth history of density perturbation can be described by EOS, $\omega$, and GIP, $\gamma$, respectively. Thus, we assume that both $\omega$ and $\gamma$ is determined from the observations. We briefly review how to reconstruct STG model functions $F(\phi)$ and $U(\phi)$ as a function of redshift, $z$ in a JF by using $\omega$ and $\gamma$ in the next section \cite{10122646}. In Sec. 3, we derive the evolution of Hubble parameter, $H$ and that of effective dark energy EOS, $\omega$ as functions of $z$ in EF. We show the discrepancies between the values of Hubble parameter in Jordan frame and those in Einstein frame. We also investigate the differences between the Jordan frame effective dark energy equation of state and the Einstein frame one.   

\section{Reconstructions of $F(\phi)$ and $U(\phi)$}
\setcounter{equation}{0}

We denote the frame dependent variables by using bar on them. Both the physical time, $\bart$ and the scale factor, $\bara$ depend on frame. However, they are not the measurable quantities in the Universe. The measurable variable is the redshift, $\barz$ and this is a frame independent value  as we show later. The action of the STG in the JF is given by \cite{0009034}
\be S_{{\rm J}} = \fr{1}{16 \pi G_{\ast}} \int d^4 x \sqrt{-\barg} \Biggl[ F(\phi) \barR - \overline{g}^{\mu\nu}\nabla_{\mu} \phi \nabla_{\nu} \phi - 2 U(\phi) \Biggr] + S_{\m}(\barg_{\mu\nu}, \psi_{\m}) \, , \label{SJF} \ee
where $G_{\ast}$ is the bare gravitational coupling constant which differs from the measured one, $\barg_{\mu\nu}$ is the JF metric, $F(\phi)$ is dimensionless function of scalar field, and $U(\phi)$ is the potential of the scalar field $\phi$. $F(\phi)$ needs to be positive to ensure that the gravity is attractive. $S_{\m}(\barg_{\mu\nu})$ is the matter action and  the matter fields $\psi_{\m}$ is universally coupled to the metric $\barg_{\mu\nu}$. All experimental data including the Hubble parameter $\barH = d \ln [\bara] / d \bart$ and redshift $\barz = \bara_{\obs} / \bara_{\s} - 1$ will thus have their usual interpretation in this JF. 

The evolution equations for the background and the matter perturbation in the flat Friedmann-Lemaitre-Robertson-Walker (FLRW) metric are given by \cite{10122646}
\ba && 3 F_{0} \barH^2  = 8 \pi G_{\ast} \barrho_{\m} + \fr{1}{2} \barH^2 \phi^{'2} - 3 \barH^2 F' + 3 \barH^2 (F_0 - F) + U  \, , \label{G00n} \\ 
&& 2 F_0 \barH \, \barH' = - 8 \pi G_{\ast} \barrho_{\m} - \barH^2 \phi^{'2} - \barH^2 F'' + (\barH^2 - \barH \, \barH') F' + 2 \barH \, \barH' (F_0 - F) \, , \label{Giin} \\ 
&& \phi'' + \Biggl( 3 + \fr{\barH'}{\barH} \Biggr) \phi' = 3 \Biggl( 2 + \fr{\barH'}{\barH} \Biggr) \fr{F'}{\phi'} - \fr{1}{\barH^2} \fr{U'}{\phi'} \, , \label{phieqn} \\ && \barrho_{\m}^{'} + 3 \barrho_{\m} = 0 \, , \label{rhoin} \\ 
&& \delta_{\m}'' + \Bigl( 2 + \fr{\barH'}{\barH}\Bigr) \delta_{\m}' - \fr{4 \pi G_{\eff} \rho_{\m}}{\barH^2} \delta_{\m} \simeq 0 \, , \, {\rm where} \, G_{\eff} = \fr{G_{\ast}}{F} \Biggl( \fr{2F + 4F_{,\phi}^2}{2F + 3F_{,\phi}^2} \Biggr)  \, ,\label{ddotdelta2} \ea 
where primes denote the differentiations with respect to the e-folding number in JF, $\barn = \ln{\bara}$, $F_{0}$ means $F(\barn=0)$, $\delta_{\m}$ is the matter density fluctuation, and we only consider the matter component which is relevant to the late-time universe. Later, we express above equations in terms of the redshift, $\barz$ which is the frame independent quantity \cite{13081142}. 

It is shown that two unknown functions of STG, $F(\phi)$ and $U(\phi)$, can be reconstructed from the combination of observations of the background evolution and the structure formation  \cite{10122646}. The background evolution and the growth history of the matter fluctuation can be parametrized by EOS $\bar{\omega}$ and GIP $\bargamma$, respectively. In other world, one can specify both $F(\phi)$ and $U(\phi)$ as long as both $\baromega$ and $\bargamma$ are well measured in observations. We adopt CPL $\baromega$ and same form of $\bargamma$ in this reconstruction. The reconstructed $F(\barn)$ and $U(\barn)$ are given by
\ba \fr{F(\barn)}{F_{0}} &=& \fr{3}{2} \fr{\Omn}{P(\barn)} \,\,\,\, , \label{FoF0n} \\ 
P(\barn) &=& \Omn^{\gamma} \Biggl( \Omn^{\gamma} + \gamma' \ln \Omn - \gamma \Bigl[ 3 + 2 \fr{\barH'}{\barH} \Bigr] + 2 + \fr{\barH'}{\barH} \Biggr) \, , \label{Pn}  \\
\fr{U(\barn)}{F_{0} \barH_{0}^2} &=& \fr{1}{2} \fr{\barH^2}{\barH_{0}^2} \Biggl( \fr{F''}{F_0} + \Bigl[ 5 + \fr{\barH'}{\barH} \Bigr] \fr{F'}{F_0} + 2 \Bigl[ 3 + \fr{\barH'}{\barH} \Bigr] \fr{F}{F_0} - 3 \Omn \Biggr) \, , \label{Un} \ea
where we should assert that $|F(\barn)| \gg |F_{, \, \phi}(\barn)^2|$. For later use, we replace above equations (\ref{FoF0n})-(\ref{Un}) as a function of the JF redshift $\barz$. 
\ba \fr{F(\barz)}{F_{0}} &=& \fr{3}{2} \fr{\Omz}{P(\barz)} \,\,\,\, , \label{FoF0z} \\ 
P(\barz) &=& \Omz^{\gamma} \Biggl( \Omz^{\gamma} -(1+\barz) \gamma_{,\barz} \ln \Omz - \gamma \Bigl[ 3 - 2 (1+\barz) \fr{\barH_{,\barz}}{\barH} \Bigr] \nonumber  \\ &&~~~~~~~~~~~+  2 - (1+\barz) \fr{\barH_{,\barz}}{\barH} \Biggr) \, , \label{Pz}  \\
\fr{U(\barz)}{F_{0} \barH_{0}^2} &=& \fr{1}{2} \fr{\barH^2}{\barH_{0}^2} \Biggl( (1+\barz)^{2} \fr{F_{,\barz\, \barz}}{F_0} + (1+\barz) \fr{F_{,\barz}}{F_0} -(1+\barz) \Bigl[ 5 -(1+\barz) \fr{\barH_{,\barz}}{\barH} \Bigr] \fr{F_{,\barz}}{F_0} \nonumber \\ 
&&~~~~~~~~~+ 2 \Bigl[ 3 -(1+\barz) \fr{\barH_{,\barz}}{\barH} \Bigr] \fr{F}{F_0} - 3 \Omz \Biggr) \, , \label{Uz} \ea 

As shown in the appendix, the redshifts in both the JF and the EF are the same and dynamics can be described as a function of $\barn = \ln \bara = - \ln (1 + \barz) = - \ln (1 +z)$. Thus, background evolution quantities are given by
\ba \Omega_{\m}(\barn) &=& \fr{\Omo}{\Omo + (1-\Omo) e^{-3(\omega_{0} + \omega_{a}) \barn} e^{-3\omega_{a} (1- e^{\barn})}} \label{Omgamz} \\
\fr{\barH^2(\barn)}{\barH_0^2} &=& \Omo e^{-3\barn} + (1 - \Omo) e^{-3(1 + \oo + \oa)\barn} e^{-3 \oa(1- e^{\barn})} \, , \label{H2oH0} \\
\fr{\barH'(\barn)}{\barH(\barn)} &=& -\fr{3}{2} \Biggl[ 1 + \omega(\barn) \fr{ (1-\Omo) e^{-3(\omega_{0} + \omega_{a})\barn} e^{-3\omega_{a}(1- e^{\barn})}}{\Omo +  (1-\Omo) e^{-3(\omega_{0} + \omega_{a})\barn} e^{-3\omega_{a}(1- e^{\barn})}} \Biggr] \, , \label{HpoH} \\
\overline{\omega}(\barn)  &=& -1 + 2 \fr{ (\fr{\phi'}{\sqrt{F}})^2 + \fr{F^{''}}{F} -\fr{F'}{F} + (\fr{F'}{F} + 2 -2 \fr{F_{0}}{F})\fr{\barH'}{\barH}}{ (\fr{\phi'}{\sqrt{F}})^2  - 6 \fr{F'}{F} + 6(\fr{F_0}{F} -1) + 2\fr{U}{F\barH^2}} \label{omde} \\
\fr{\phi'}{\sqrt{F}} &=& \sqrt{-\fr{F^{''}}{F} + \Bigl(1-\fr{\barH'}{\barH} \Bigr) \fr{F'}{F} - 2\fr{\barH'}{\barH} - 3 \fr{F_{0}}{F} \Omz } \label{phiz} \\
\fr{F'}{F_0} &=& - \Biggl( 3 + 2 \fr{\barH'}{\barH} + \fr{P'}{P} \Biggr) \fr{F}{F_{0}} \, , \label{FpoF0z} \\ 
\fr{F''}{F_0} &=& \Biggl(  \Bigl[3 + 2 \fr{\barH'}{\barH}  + \fr{P'}{P} \Bigr]^2 -2 \Bigl(\fr{\barH'}{\barH} \Bigr)' - \Bigl(\fr{P'}{P}\Bigr)' \Biggr) \fr{F}{F_{0}} \label{FppoF0z}  \ea 
where $\barH_{0} \equiv \barH (\barn = 0)$.

\section{Conformal Transformation}
\setcounter{equation}{0}
We do the conformal transformation, $\bar{g}_{\mu\nu}=F^{-1}(\phi)g_{\mu\nu}$, of the JF action in Eq.(\ref{SJF}) to obtain 
\begin{align}\label{SEF}
S_{{\rm E}} = \fr{1}{16 \pi G_{\ast}} \int d^4 x \sqrt{-g} \Biggl[ R - 2g^{\mu\nu}\nabla_{\mu} \varphi \nabla_{\mu} \varphi - 4 V(\varphi) \Biggr] + S_{\m}(g_{\mu\nu}, \psi_{\m}) \, .
\end{align}
From the above Eq.(\ref{SEF}), one obtains the EF Friedmann equations, 
\ba 3 H^{2} &\equiv& \Bigl( \fr{1}{a} \fr{da}{dt} \Bigr)^2  = 8 \pi G_{\ast} \rho + \Bigl( \fr{d \varphi}{dt} \Bigr)^2 + 2 V(\varphi) \label{G00EF} \\
-\fr{3}{a} \fr{d^2 a}{dt^2} &=& 4 \pi G_{\ast} (\rho + 3P) + 2 \Bigl( \fr{d \varphi}{dt} \Bigr)^2 - 2 V(\varphi) \label{GiiEF} \, , \ea
where the scale factor $a \equiv \sqrt{F} \bar{a}$ and the physical time $dt \equiv \sqrt{F} d\bar{t}$ defined in the EF. Even though the above Eqs.(\ref{G00EF})-(\ref{GiiEF}) are simple, they are not the observed quantities \cite{13081142}. Instead one needs to express the above equations in terms of observable quantities which is frame-independent, e.g., $z = \barz = 1/\bar{a} -1$ \cite{13081142}. The relation between the Hubble parameter in the JF and one in the EF, $\overline{H}$ and $H$, are related to each other by
\begin{align}
H &\equiv \frac{1}{a}\frac{d a}{d t}= \fr{\barH}{\sqrt{F}} \Bigl( \fr{F'+2F}{2F} \Bigr) \equiv \fr{\barH}{\sqrt{F}} B^{-1}  \equiv \HE B^{-1} \, \rightarrow \, \HE \equiv B H = \fr{\barH}{\sqrt{F}} \label{HE} \\
\fr{1}{\HE} \fr{d \HE}{d \ln a} &= 
B \Bigl( \fr{\barH'}{\barH} - \fr{1}{2} \fr{F'}{F} \Bigr) \label{HEp}
\end{align}
where the conversion factor $B$ from $H$ to the observed EF Hubble parameter $H_{{\rm E}}$ is defined by
$\displaystyle B \equiv \fr{ d \ln \bara}{d \ln a} = \fr{2F}{2F + F'}$\cite{13081142}. 
In terms of the observed Hubble parameter $H_{{\rm E}}$, we write the EF Friedmann equations again as
\ba 3 H_{\e}^2 &=& 8 \pi G_{\ast} \rho_{\m} + \Bigl[ \fr{1}{2} \Bigl(\fr{\phi'}{\sqrt{F}}\Bigr)^2 - 3 \fr{F'}{F} + \fr{U}{F \barH^2} \Bigr] H_{\e}^2 \equiv 8 \pi G_{\ast} (\rho_{\m} + \rho_{\DE}) \, ,  \label{G00E} \\
-2 H_{\e} \fr{d H_{\e}}{d \ln a} &=& 8 \pi G_{\ast} \rho_{\m} + \Bigl[ \Bigl(\fr{\phi'}{\sqrt{F}}\Bigr)^2 + \fr{1}{2} \Bigl(\fr{F'}{F} \Bigr)^2 + \fr{F''}{F} \Bigr] H_{\e}^2 \equiv 8 \pi G_{\ast} \Bigl(\rho_{\m} + (1+\omega_{\e}) \rho_{\DE} \Bigr) \, . \label{GiiE} \ea
From the above Eqs.(\ref{G00E}) and (\ref{GiiE}), one can define the observed effective DE EOS in the EF $\wE$ by
\be \wE = -1 - \fr{2}{3} B \Bigl( \fr{\barH'}{\barH} - \fr{1}{2} \fr{F'}{F} \Bigr) 
= -1 + 2 \fr{(\fr{\phi'}{\sqrt{F}})^2 + \fr{F''}{F} + \fr{1}{2} (\fr{F'}{F})^2}{(\fr{\phi'}{\sqrt{F}})^2 -6 \fr{F'}{F} + 2\fr{U}{F\barH^2}}\,. \label{omegaE}  \ee

One can also obtain the model functions $( \fr{d \varphi}{dt} )^2$ and $V(\varphi)$ as 
\ba \Bigl( \fr{\grave{\varphi}}{B} \Bigr)^2 &=& \fr{1}{2} \Bigl( \fr{\phi'}{\sqrt{F}} \Bigr)^2 + \fr{3}{4} \Bigl( \fr{\grave{F}}{B F} \Bigr)^2 \, , \label{dvarphi} \\
2V(\varphi) &=& \fr{U(\phi)}{F(\phi)^2} \label{Vvarphi} \, , \ea
where $\grave{\varphi} \equiv d \varphi / d \ln a$ and $\grave{F} \equiv d F / d \ln a$. 
Because both $F(\phi)$ and $U(\phi)$ are obtained from observations, one can obtain the evolutions of $\varphi$ and $V$ from observations. We summarize the frame dependent EOSs of the effective dark energy in terms of $\ln \bara$ and $\ln a$ in table.\ref{table:eos}. Primes mean the derivatives with respect to the $\ln \bara$ and graves denote the derivative with respect to the $\ln a$.    

\begin{table}
\centering
\begin{tabular}{c cc}
\hline \hline
                & $\ln \bara$   & $\ln a$ \\            
\hline
\\[0.2ex]
$1 + \overline{\omega}_{\rm{DE}}$ & $\fr{(\fr{\phi'}{\sqrt{F}})^2 + \fr{F''}{F} - \fr{F'}{F} + (\fr{F'}{F} + 2 - 2\fr{F_{0}}{F}) \fr{\barH'}{\barH}}{\fr{1}{2}(\fr{\phi'}{\sqrt{F}})^2 -3 \fr{F'}{F} + \fr{U}{F\barH^2} + 3(\fr{F_{0}}{F}-1)}$ & $\fr{2 \Bigl(\fr{\grave{\varphi}}{B} \Bigr)^2 + \Bigl(\fr{2}{B^2} - \fr{F_0}{F} \fr{1}{B} \Bigr)  \fr{\grave{H}_{\e}}{H_{\e}} 
+ \fr{1}{B^3} \Bigl[ \fr{\grave{\grave{F}}}{F} - \fr{1}{2} \Bigl(\fr{\grave{F}}{F} \Bigr)^{3} - B \Bigl(\fr{\grave{F}}{F} \Bigr)^2 \Bigr] - \fr{1}{2B} \fr{F_0}{F} \fr{\grave{F}}{F} }{\Bigl(\fr{\grave{\varphi}}{B} \Bigr)^2 +  \fr{2V}{H_{\e}^2} + 3(\fr{F_{0}}{F}-1) - \fr{3}{B} \fr{\grave{F}}{F} \Bigl( 1 + \fr{1}{4B} \fr{\grave{F}}{F} \Bigr) }$\\[5ex]
\hdashline
\\[0.2ex]
$1 + \omega_{\rm{DE}}^{(\rm{E})}$ &     $\fr{(\fr{\phi'}{\sqrt{F}})^2 + \fr{F''}{F} + \fr{1}{2} (\fr{F'}{F})^2}{\fr{1}{2}(\fr{\phi'}{\sqrt{F}})^2 -3 \fr{F'}{F} + \fr{U}{F\barH^2}}$    &  $\fr{2 \Bigl(\fr{\grave{\varphi}}{B} \Bigr)^2 + \fr{1}{B^3} \Bigl[ \fr{\grave{\grave{F}}}{F} - \fr{1}{2} \Bigl( \fr{\grave{F}}{F} \Bigr)^{3} - B \Bigl( \fr{\grave{F}}{F} \Bigr)^{2} \Bigr]}{ \Bigl(\fr{\grave{\varphi}}{B} \Bigr)^2 + \fr{2V}{H_{\e}^2} - \fr{3}{B} \fr{\grave{F}}{F} \Bigl[ \fr{4- \fr{\grave{F}}{F}}{4- 2\fr{\grave{F}}{F}} \Bigr]}$ \\[5ex]
\hline
\end{tabular}
\caption{EOSs of the effective dark energy in JF and  EF.} 
\label{table:eos}
\end{table}

\begin{figure}[h]
\centering
\vspace{1cm}
\begin{tabular}{cc}
\epsfig{file=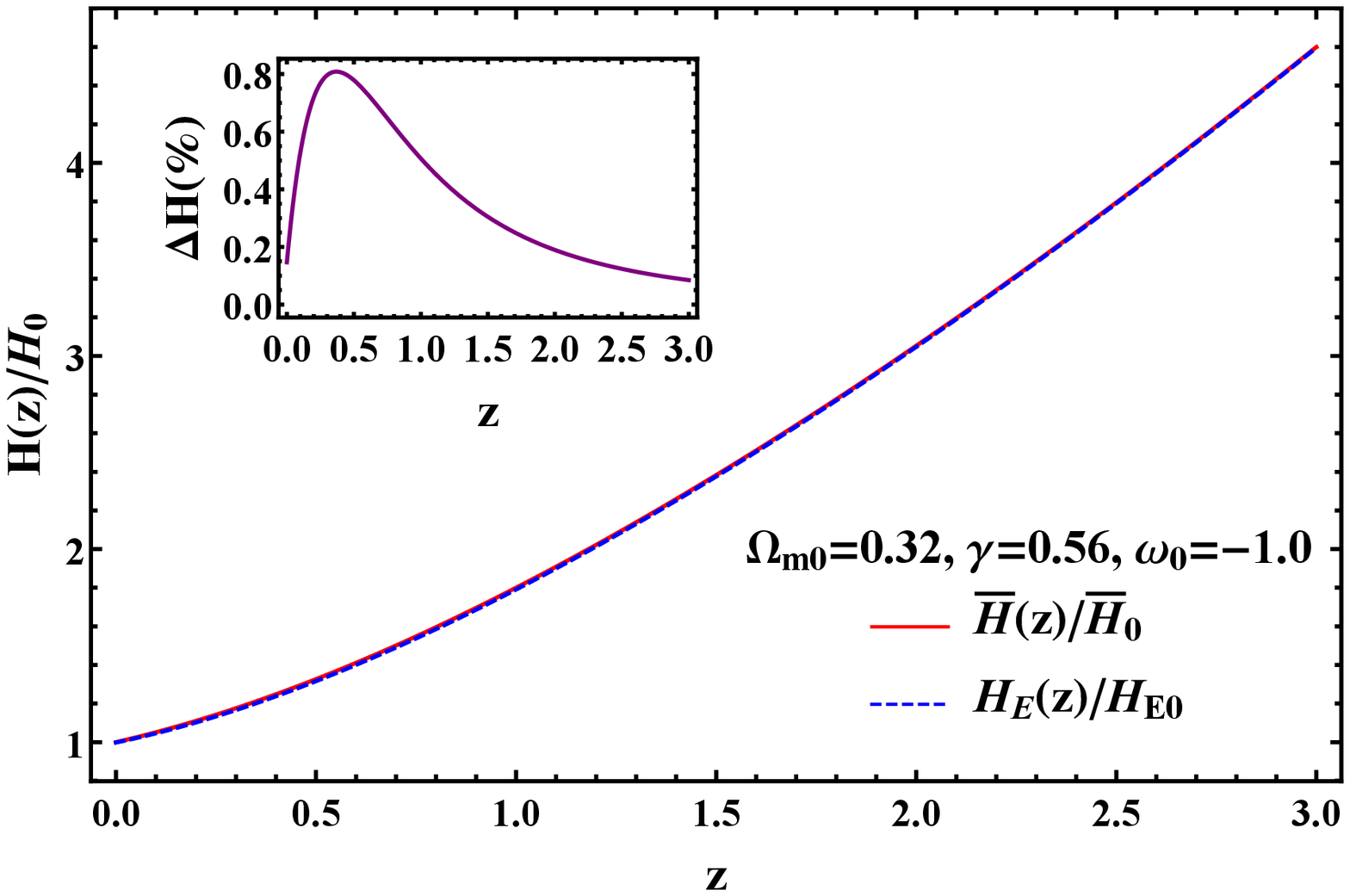,width=0.45\linewidth,clip=} &
\epsfig{file=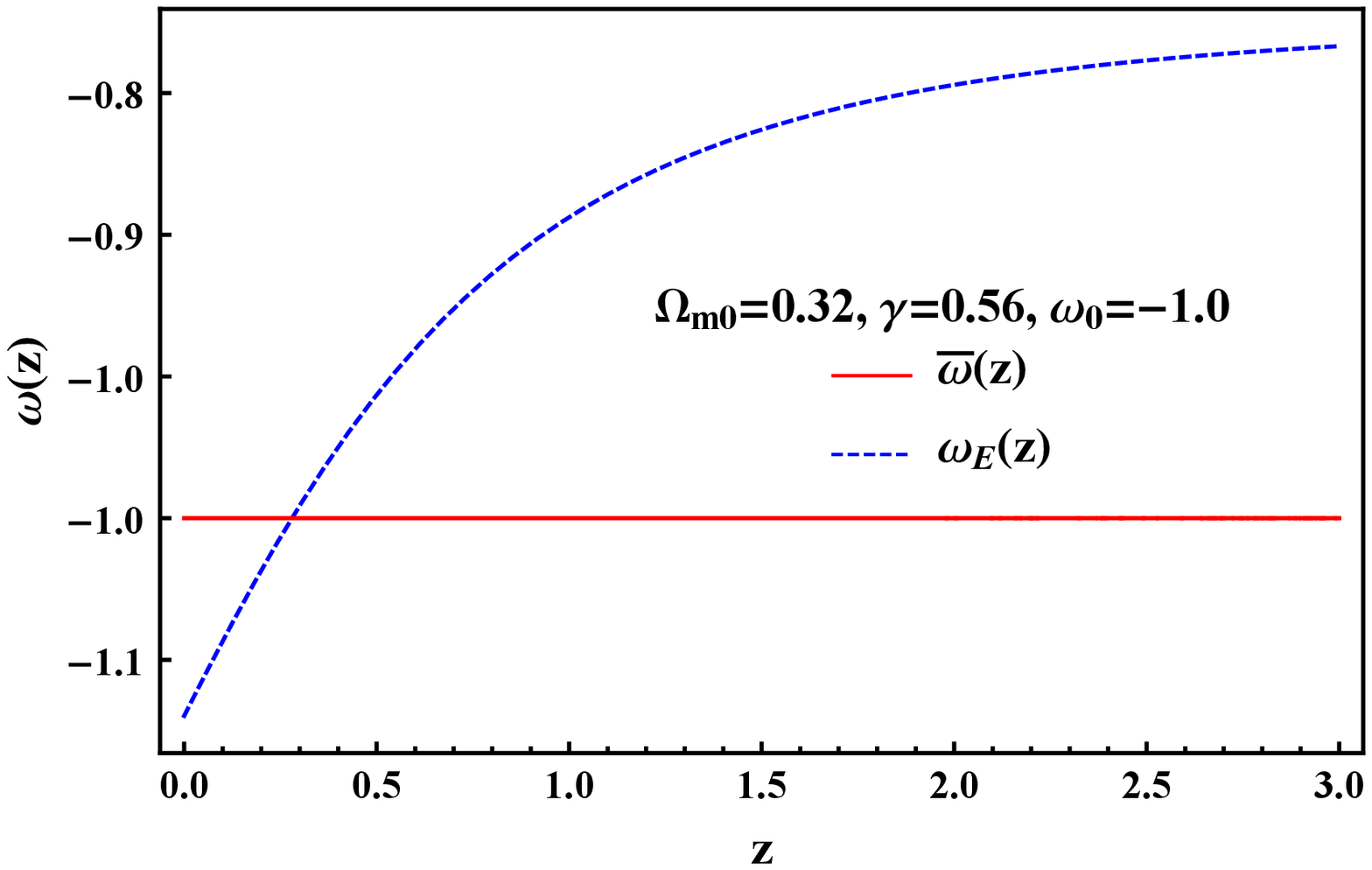,width=0.467\linewidth,clip=} 
\end{tabular}
\vspace{-0.5cm}
\caption{ a) Evolutions of the normalized Hubble parameters both on Jordan frame (solid line) and on Einstein frame (dashed line) when $\Omega_{\m 0} =0.32, \omega_{0} = -1.0$, and $\gamma_{0} = 0.56$. b) Evolution of effective equation of states of the dark energy for both frames.} \label{fig1}
\vspace{1cm}
\end{figure}

One can compare the evolutions of Hubble parameters both in the JF and in the EF. As given in Eq.(\ref{HE}), the difference of Hubble parameters in both frames is given by $\sqrt{F}$. Thus, based on the solar system test, one expects this discrepancy of Hubble parameters in two frames should be quite small. This is shown in the left panel of Fig.\ref{fig1}. We adopt $\Lambda$CDM model with $\Omo = 0.32$ for the background evolution, $\overline{\omega} = -1.0$ and the GIP is also same as that of $\Lambda$CDM, $\gamma = 0.56$. Thus, in this model, one will not be able to distinguish the $\Lambda$CDM from the STG. However, if one can measure the accurate Integrated Sachs-Wolfe (ISW) effect, then it is still possible to distinguish between two models \cite{10122646}. The solid line depicts the evolution of the Hubble parameter normalized by its present values in the JF, $\barH / \barH_{0}$. The evolution of the EF Hubble parameter normalized by its present value, $\HE / H_{\rm{E}0}$, is illustrated by the dashed line. The difference between the JF Hubble parameter and that of the EF, $\fr{(\barH/\barH_{0} - \HE / H_{\rm{E}0})}{\barH/\barH_{0}} \times 100$, is less than 1 \% for all redshifts as shown in the left panel of Fig.\ref{fig1}. Thus, one needs sub-percent level accuracies of the Hubble parameter measurement in order to distinguish the physical frame from observations. Now, one can compare the effective DE EOSs both in the JF, $\overline{\omega}$, and in the EF, $\omega_{\rm{E}}$. As required, the JF DE EOS is consistent with that of $\Lambda$ as $-1.0$ and this is shown in the right panel of Fig.(\ref{fig1}). The solid line in that panel depicts the JF DE EOS. The EF DE EOS is described by the dashed line and this is quite different from that of the JF. The percentage difference of the DE EOS between the two frames is bigger than 7 \% at all redshifts as shown in the right panel of Fig.\ref{fig1}. Thus, if one interpret the cosmological parameters in the JF, then one should use the JF as a physical frame. Due to the matter coupling in the EF, it is not conformally invariant for the observables. 

\begin{figure}[h]
\centering
\vspace{1cm}
\begin{tabular}{cc}
\epsfig{file=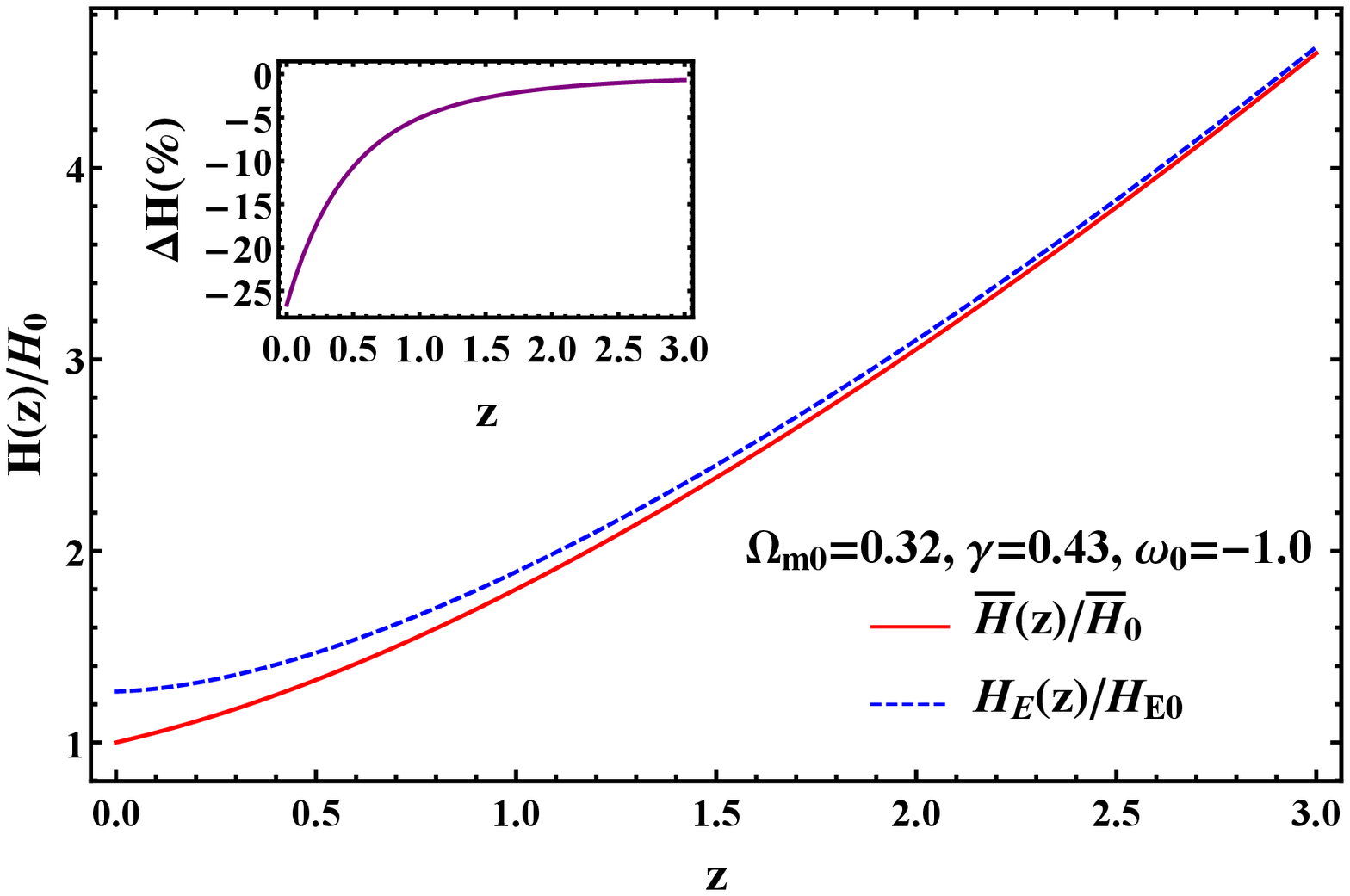,width=0.45\linewidth,clip=} &
\epsfig{file=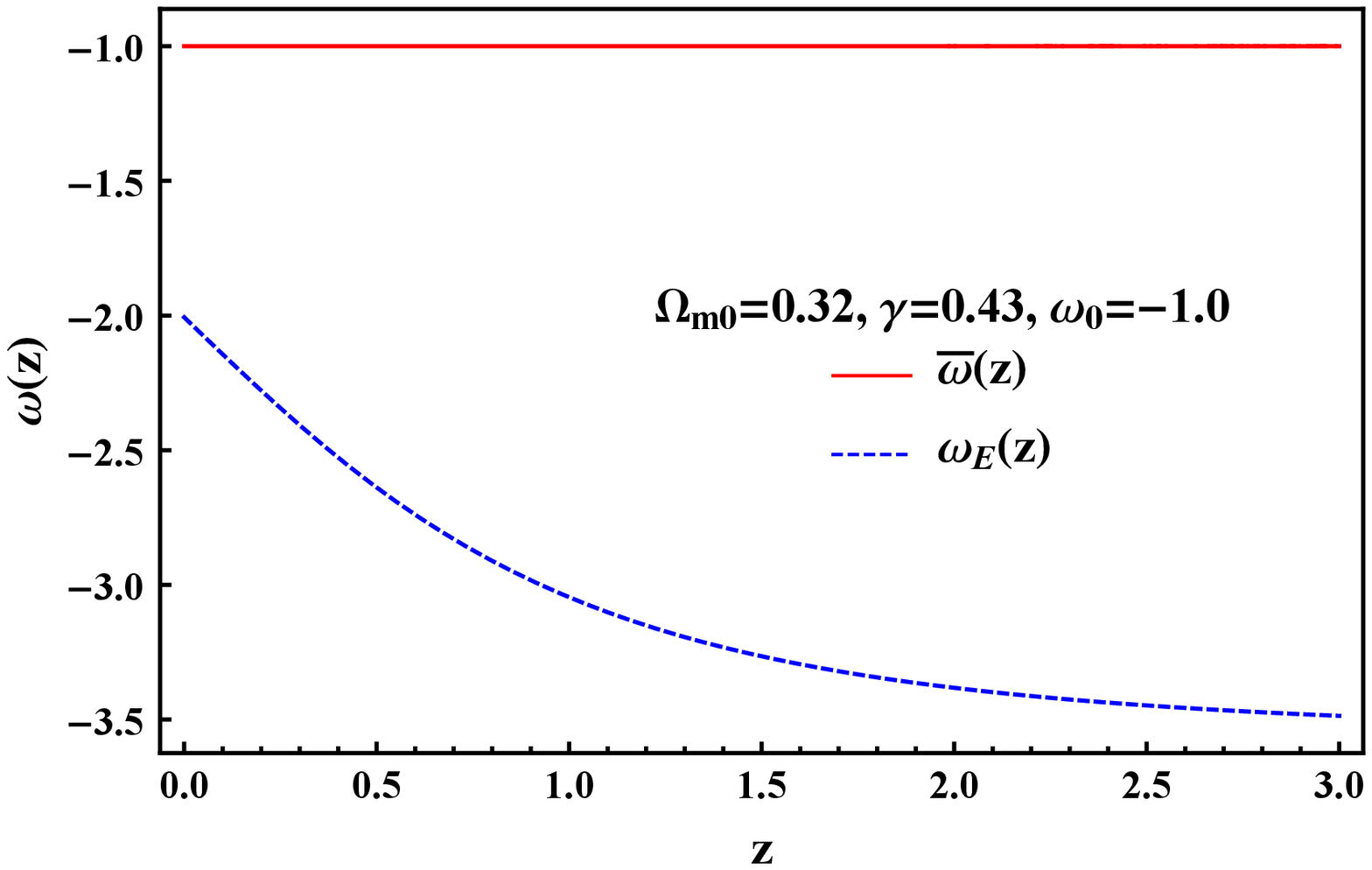,width=0.467\linewidth,clip=} 
\end{tabular}
\vspace{-0.5cm}
\caption{ a) Evolutions of the normalized Hubble parameters both on Jordan frame (solid line) and on Einstein frame (dashed line) when $\Omega_{\m 0} =0.32, \omega_{0} = -1.0$, and $\gamma_{0} = 0.43$. b) Evolution of effective equation of states of the dark energy for both frames.} \label{fig2}
\end{figure}

\begin{figure}[h]
\centering
\vspace{1cm}
\begin{tabular}{cc}
\epsfig{file=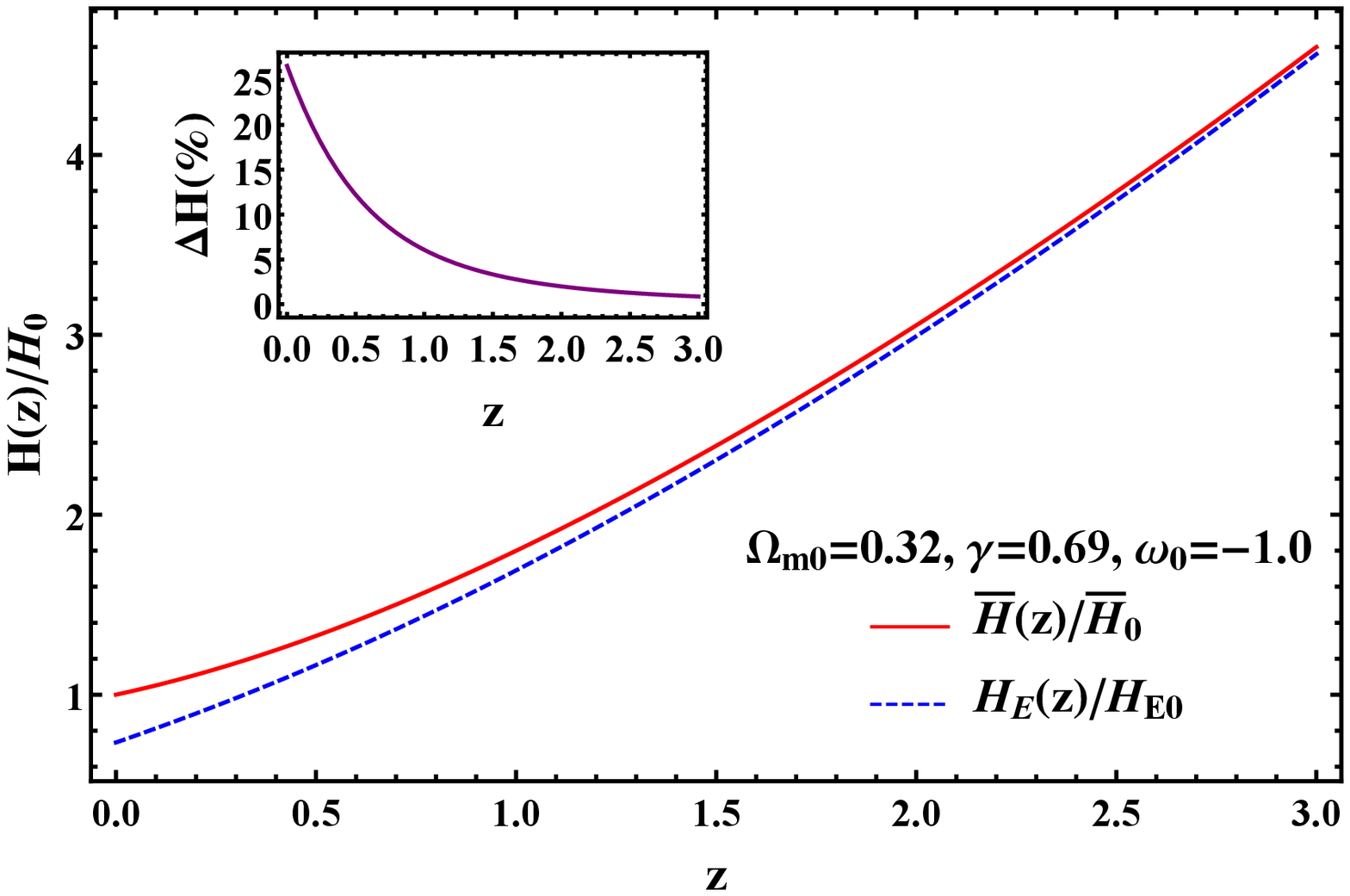,width=0.45\linewidth,clip=} &
\epsfig{file=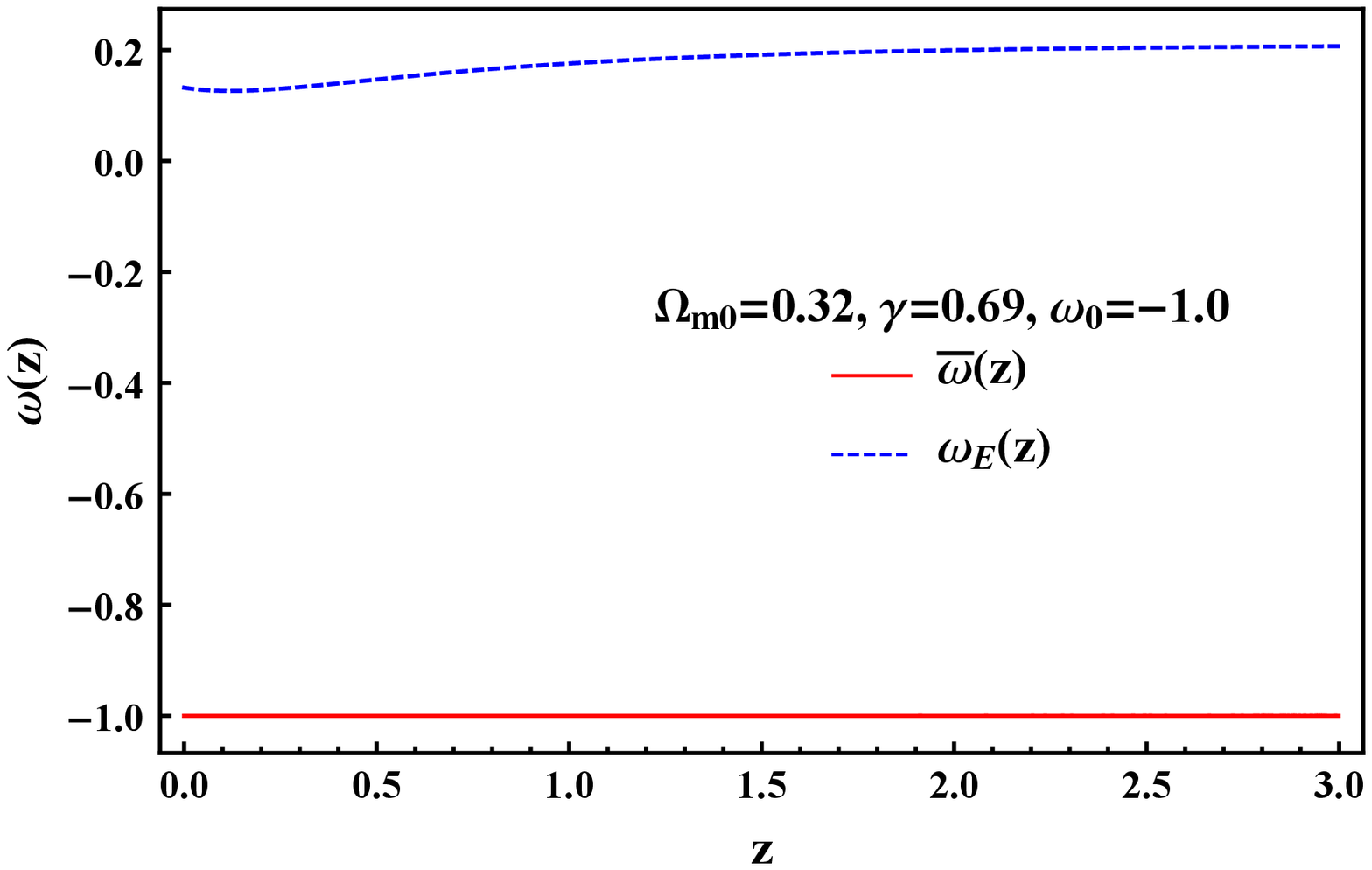,width=0.467\linewidth,clip=} 
\end{tabular}
\vspace{-0.5cm}
\caption{ a) Evolutions of the normalized Hubble parameters both on Jordan frame (solid line) and on Einstein frame (dashed line) when $\Omega_{\m 0} =0.32, \omega_{0} = -1.0$, and $\gamma_{0} = 0.69$. b) Evolution of effective equation of states of the dark energy for both frames.} \label{fig3}
\vspace{1cm}
\end{figure}

In Fig.\ref{fig2} and Fig.\ref{fig3}, we show different behaviors of  the normalized Hubble parameters and the DE EOSs for the different values of $\gamma_{0}$ when $\Omega_{\m 0} =0.32, \omega_{0} = -1.0$. This difference is occurred if we assume that the observational measurement error in $f$-value is 10 \% \cite{170508797}. If $f$ measurement has  $+10\,$\% error, then $\gamma_0 = 0.43$ as shown in Fig.\ref{fig2}. For $-10\,$\% error, $\gamma_0 = 0.69$ as in Fig.\ref{fig3}.

\begin{figure}[h]
\centering
\vspace{1cm}
\begin{tabular}{cc}
\epsfig{file=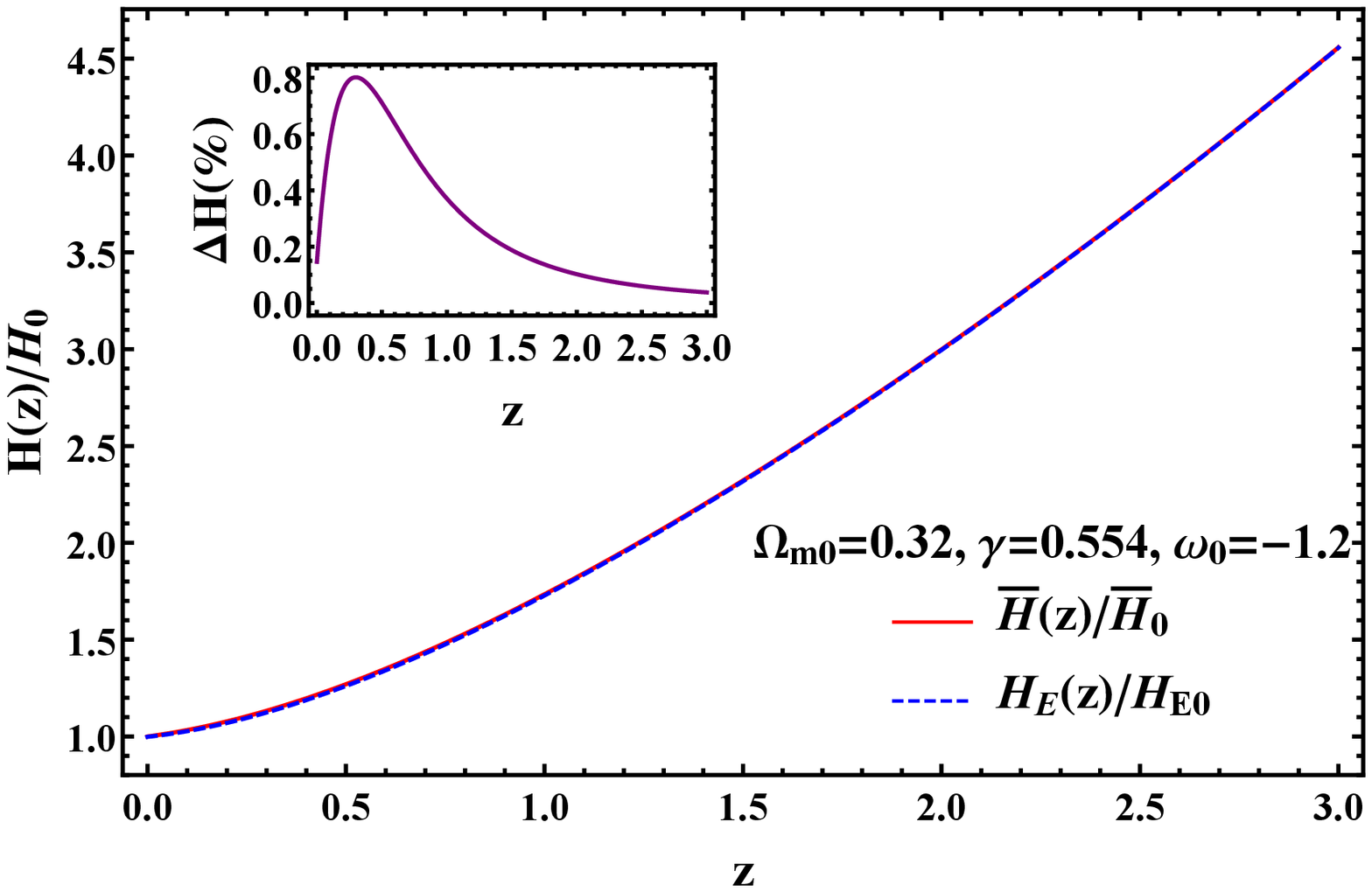,width=0.45\linewidth,clip=} &
\epsfig{file=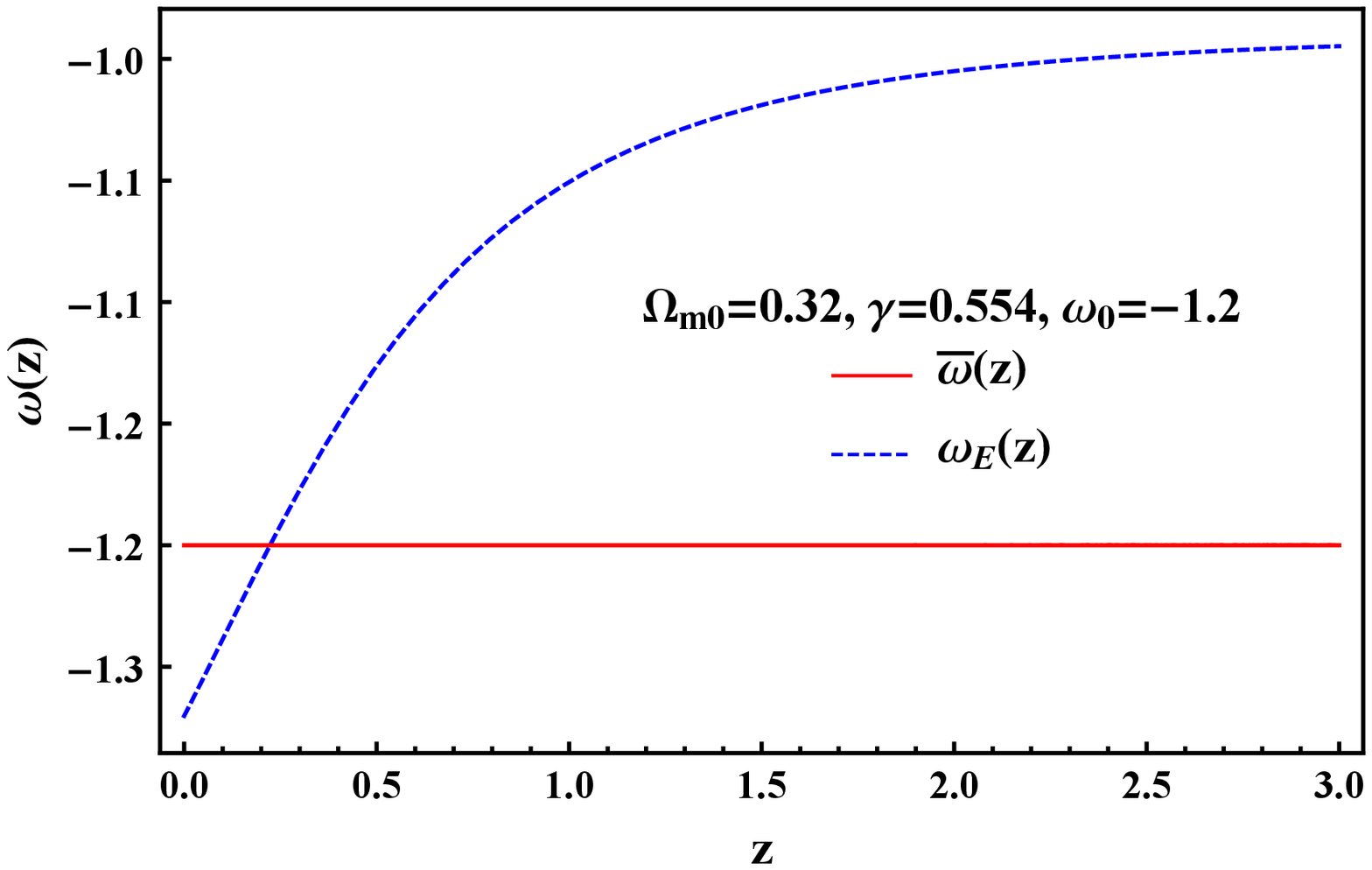,width=0.467\linewidth,clip=} 
\end{tabular}
\vspace{-0.5cm}
\caption{ a) Evolutions of the normalized Hubble parameters both on Jordan frame (solid line) and on Einstein frame (dashed line) when $\Omega_{\m 0} =0.32, \omega_{0} = -1.2$, and $\gamma_{0} = 0.554$.  b) Evolution of the effective equation of states of the dark energy for both frames.} \label{fig4}
\end{figure}

We investigate other models. We probes $\Omega_{\m 0} =0.32, \omega_{0} = -1.2$, and $\gamma_{0} = 0.554$ model in Fig. \ref{fig4}. In this model, Hubble parameters are almost identical to each other in both frames. The Hubble parameters in JF and in EF described by solid and dashed lines, respectively. This is shown in the left panel of Fig.\ref{fig4}. However, the EOS in the EF deviates from that of the JF as much as 17 \%. Also, its early time behavior mimics that of $\Lambda$ until $z > 0.5$. This is shown as the dashed line in the right panel of Fig.\ref{fig4}.    

\begin{figure}[h]
\centering
\vspace{1cm}
\begin{tabular}{cc}
\epsfig{file=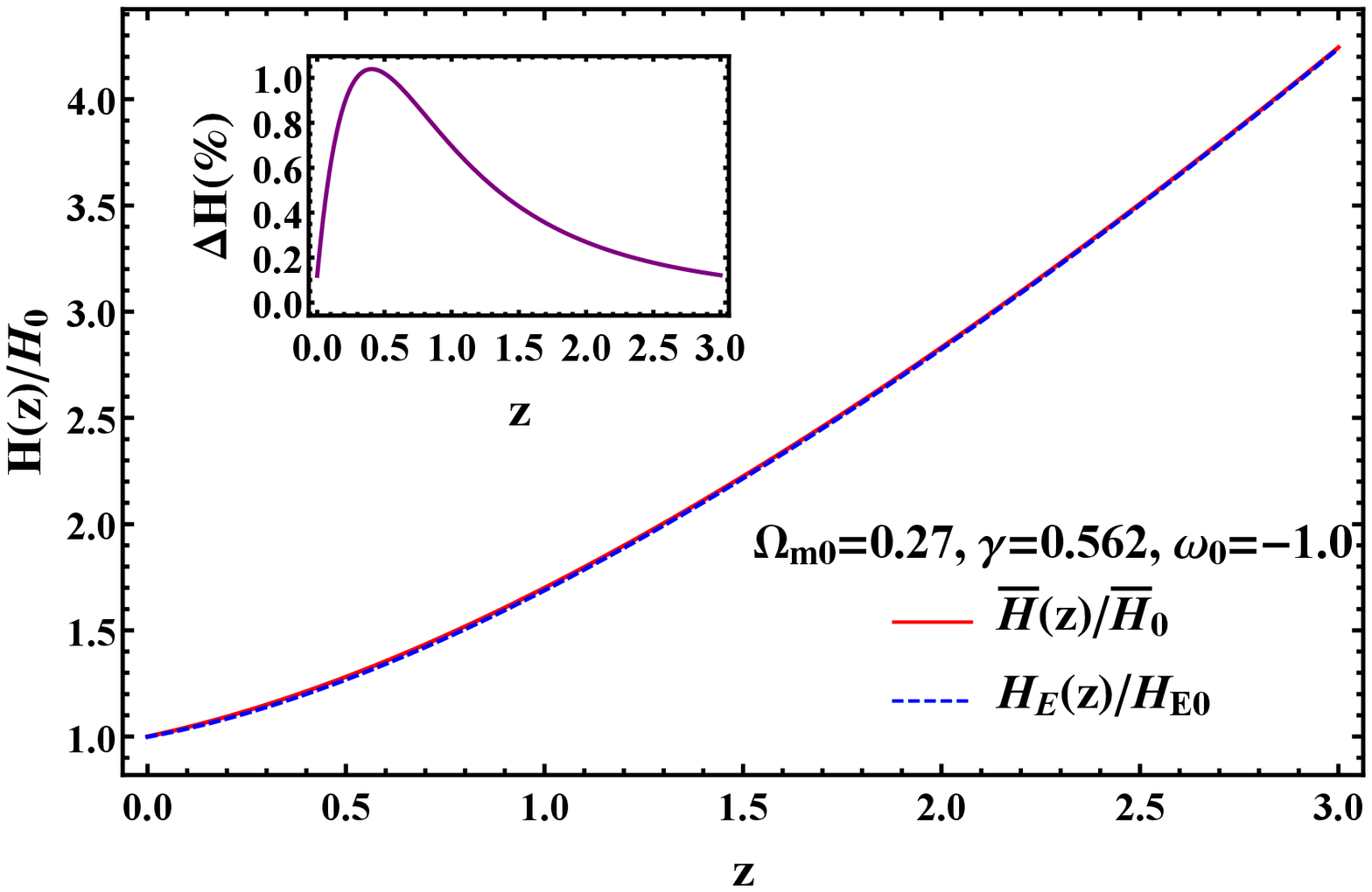,width=0.45\linewidth,clip=} &
\epsfig{file=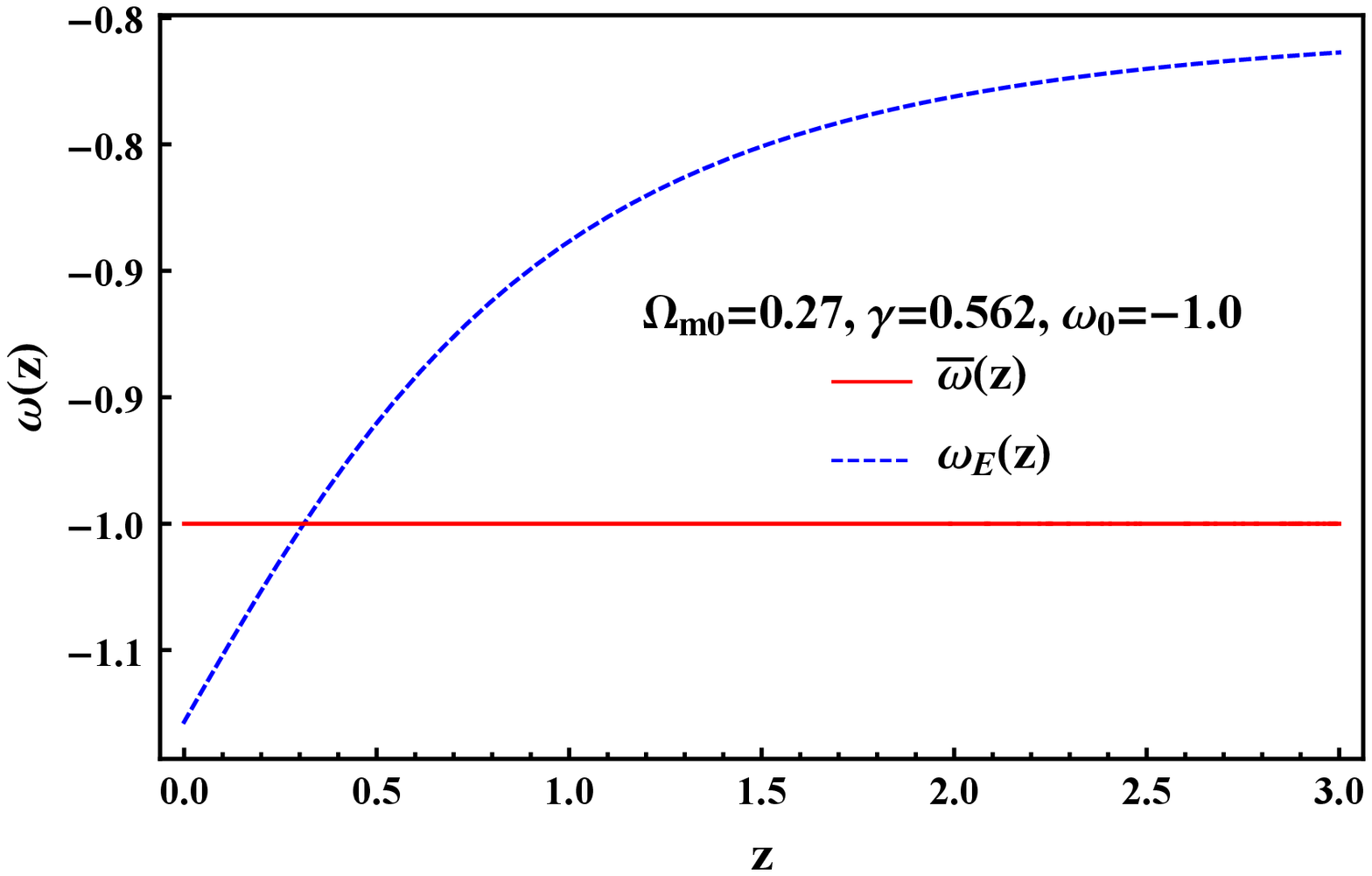,width=0.467\linewidth,clip=} 
\end{tabular}
\vspace{-0.5cm}
\caption{ a) Evolutions of the Hubble parameters normalized by its present values both on Jordan frame (solid line) and on Einstein frame (dashed line) when $\Omega_{\m 0} =0.27, \omega_{0} = -1.0$, and $\gamma_{0} = 0.562$. b) Evolution of effective equation of motion of dark energy in both frames for same models.} \label{fig5}
\vspace{1cm}
\end{figure}

We also probe $\Omega_{\m 0} =0.27, \omega_{0} = -1.0$, and $\gamma_{0} = 0.562$ model in Fig.\ref{fig5}. The normalized Hubble parameters in the JF and in the EF described by solid and dashed lines, respectively. This is shown in the left panel of Fig.\ref{fig5}. We also show the effective DE EOS for this model in the right panel of Fig.\ref{fig5}. As in the first model, this model can describe the phantom crossing in the EF.    

\section*{Conclusions}
We show that unknown model functions of the scalar tensor gravity theories are determined from cosmological observations. Thus, if cosmological observables are accurate enough, then one can constrain the model functions. As we show even if the future observations are consistent with $\Lambda$CDM in both background evolutions and the growth rate of the large scale structure, the integrated Sachs-Wolfe effect can distinguish between $\Lambda$CDM and scalar tensor gravity theories. Furthermore, if one uses the one frame to obtain the model functions, then one should use the same frame in order to interpret other observables. We show the frame dependence of cosmological observables explicitly. Also, one can investigate the scalar tensor gravity models without using specific forms of them. Also, if one release the constraint of solar system test, then one can investigate more general scalar tensor gravity models which differ from general relativity  even in the Jordan frame.

\appendix
\section{Appendix}
\setcounter{equation}{0}
In this appendix, we summarize the expressions of some quantities in terms of both $\ln \bara$ and $\ln a$.

\ba \fr{d \ln a}{d \ln \bara} &=& \fr{d \ln (\sqrt{F} \bara)}{d \ln \bara} \equiv \fr{1}{2} \fr{F'}{F} + 1 = \fr{F' + 2F}{2F} \label{Binv} \\
\fr{d \ln \bara}{d \ln a} &=& \fr{d \ln (a/\sqrt{F})}{d \ln a} = 1 - \fr{1}{2} \fr{\grave{F}}{F} = \fr{2F - \grave{F}}{2F}  \label{B2} \\
B &\equiv& \fr{d \ln \bara}{d \ln a} = \fr{2F}{2F + F'} = \fr{2F - \grave{F}}{2F} \label{B} \\
\fr{F'}{F} &\equiv& \fr{d \ln F}{d \ln \bara} = \fr{d \ln a}{d \ln \bara} \fr{d \ln F}{d \ln a} \equiv \fr{1}{B} \fr{\grave{F}}{F}= \fr{2 \grave{F}}{2F - \grave{F}} \label{Fp} \\ 
\fr{F''}{F} &=& \fr{1}{B^3} \Bigl[ \fr{\grave{\grave{F}}}{F} - \fr{1}{2} \Bigl( \fr{\grave{F}}{F} \Bigr)^{3} \Bigr] \,\, , \rm{where} \,\, \grave{\grave{F}} \equiv \fr{d^2 F}{d \ln a^2}  \label{Fpp}  \\
\fr{1}{2} \Bigl(\fr{\phi'}{\sqrt{F}}\Bigr)^{2} &=& \Bigl( \fr{\grave{\varphi}}{B} \Bigr)^{2} - \fr{3}{4} \Bigl(\fr{\grave{F}}{BF} \Bigr)^2 = \Bigl( \fr{\grave{\varphi}}{B} \Bigr)^{2} - 3 \Bigl( \fr{\grave{F}}{2F - \grave{F}}  \Bigr)^{2} \label{phip2} \\
\fr{U}{F \barH^2} &=& \fr{2V}{H_{\e}^2} \label{UV} \\
1 + \barz_{\s} &\equiv& \fr{\bara_{\obs}}{\bara_{\s}} = \sqrt{\fr{F_{\s}}{F_{\obs}}} \fr{a_{\obs}}{a_{\s}} = 1 + z_{\s} \neq \fr{a_{\obs}}{a_{\s}}  \label{barzvsz} \\ 
1 + \barz &=& \fr{1}{\bara} = 1+z\,\, , \text{\rm{when}} \,\, \bara_{\obs} = 1\label{barz} \\
1 + z &=& \fr{\sqrt{F}}{a} \,\, , \text{\rm{when}} \,\, a_{\obs} = \sqrt{F_{\obs}} = 1 \label{z} \ea

It is also useful to convert the derivatives with respect to the redshift $\barz$. 
\ba F' &\equiv& \fr{d F}{d \ln \bara} =  \fr{d \barz}{d \ln \bara} \fr{d F}{d \barz} \equiv -(1+\barz) F_{,\barz} \label{Fpbarz} \\
                &=& \fr{d \ln a}{d \ln \bara} \fr{dz}{d \ln a} \fr{d F}{dz} = B^{-1} \Bigl(  \fr{1}{2} F_{,z} - \fr{1}{1+z}  \Bigr)^{-1} F_{,z} = -(1+z) F_{,z} \label{Fpz} \\
B &=& \fr{d \ln \bara}{d \ln a} = \fr{ d \ln \bara}{dz} \fr{dz}{d \ln a} = -\fr{1}{(1+z)} \Bigl(  \fr{1}{2} F_{,z} - \fr{1}{1+z}  \Bigr)^{-1} \label{Bz} \\
\fr{d \ln a}{dz} &=& \fr{d \ln (\sqrt{F}/(1+z))}{dz} = \fr{1}{2} \fr{d F}{dz} - \fr{d \ln (1+z)}{dz} = \fr{1}{2} F_{,z} - \fr{1}{1+z} \label{dlnadz} \\               
F'' &=& (1+\barz)^2 F_{,\barz \, \barz} + (1+\barz) F_{,\barz} = (1+z)^2 F_{,z \, z} + (1+z) F_{,z}  \label{Fppbarz} \\
\phi' &=& -(1+\barz) \phi_{,\barz} \label{phiz} \\
\barH' &=& -(1+\barz) \barH_{,\barz} \label{barHz} \ea

\section*{Acknowledgments}
SL is supported by Basic Science Research Program through the National Research Foundation of Korea (NRF)
funded by the Ministry of Science, ICT and Future Planning (Grant No. NRF-2015R1A2A2A01004532) and (NRF-2017R1A2B4011168). YK is supported by Basic Science Research Program through the National Research Foundation of Korea (NRF) funded by the Ministry of Science, ICT and Future Planning (Grant No. NRF-2015R1A2A2A01004532) and (NRF-2016R1D1A1B03931090).


\begin{thebibliography}{99}


\bibitem{08014606} S.~Lee, Mod.\ Phys.\ Lett.\ A {\bf 23}, 1388 (2008) [arXiv:0801.4606]. 

\bibitem{0009008} M.~Chevallier and D.~Polarski, Int.\ J.\ Mod.\ Phys.\ D {\bf 10}, 213 (2001) [arXiv:gr-qc/0009008].

\bibitem{0208512} E.~V.~Linder, \PRL {\bf 90}, 091301 (2003) [arXiv:astro-ph/0208512].


\bibitem{0507184} M.~Ishak, A.~Upadhye, D.~N.~Spergel, Phys.\ Rev.\ D {\bf 74}, 043513 (2006) [arXiv:astro-ph/0507184]. 

\bibitem{0612452} M.~Kunz and D.~Sapone, Phys.\ Rev.\ Lett.\ {\bf 98}, 121301 (2007) [arXiv:astro-ph/0612452].

\bibitem{07042421} L.~Amendola, M.~Kunz, and D.~Sapone, \JCAP {\bf 0804}, 013 (2008) [arXiv:0704.2421].

\bibitem{07090307} I.~Laszlo and R.~Bean, Phys.\ Rev.\ D {\bf 77}, 024048 (2008) [arXiv:0709.0307].

\bibitem{08012431} E.~Bertschinger and P.~Zukin, Phys.\ Rev.\ D {\bf 78}, 024015 (2008) [arXiv:0801.2431].

\bibitem{08021068} S.~F.~Daniel, R.~R.~Caldwell, A.~Cooray, and A.~Melchiorri,
Phys.\ Rev.\ D {77}, 103513 (2008) [arXiv:0802.1068].

\bibitem{08033292} H.~Wei and S.~N.~Zhang, Phys.\ Rev.\ D {\bf 78}, 023011 (2008) [arXiv:0803.3292].

\bibitem{10122646} S.~Lee, \JCAP {\bf 1103}, 021 (2011) [arXiv:1012.2646].

\bibitem{13076619} S.~Lee, \JCAP {\bf 1402}, 021 (2014) [arXiv:1307.6619].

\bibitem{160501644} S.~Fay, \MNRAS {\bf 460}, 1863 (2016) [arXiv:1605.01644]. 


\bibitem{170508797} S.~Basilakos and S.~Nesseris [arXiv:1705.08797].

\bibitem{0005016} G.~Dvali, G.~Gabadadze, and M.~Porrati, Phys.\ Lett.\ B {\bf 485}, 208 (2000) [arXiv:hep-th/0005016]. 

\bibitem{0010186} C.~Deffayet, Phys.\ Lett.\ B {\bf 502}, 199 (2001) [arXiv:hep-th/0010186].

\bibitem{0105068} C.~Deffayet, G.~Dvali, and G.~Gabadadze, Phys.\ Rev.\ D {\bf 65}, 044023 (2002) [arXiv:astro-ph/0105068].


\bibitem{0401515} A.~Lue, R.~Scoccimarro, and G.~D.~Starkman, Phys.\ Rev.\ D {\bf 69}, 124015 (2004) [arXiv:astro-ph/0401515].

\bibitem{0701317} E.~V.~Linder and R.~N.~Cahn, Astropart.\ Phys.\ {\bf 28}, 481 (2007) [arXiv:astro-ph/0701317].

\bibitem{08024122} H.~Wei, Phys.\ Lett.\ B {\bf 664}, 1 (2008) [arXiv:0802.4122].

\bibitem{09051735} X.~Fu, P.~Wu, and H.~Yu, Phys.\ Lett.\ B {\bf 677}, 12 (2009) [arXiv:0905.1735].

\bibitem{09053444} P.~Wu, H.~Yu, and X.~Fu, \JCAP {\bf 0906}, 019 (2009) [arXiv:0905.3444].


\bibitem{Buchdahl} H.~A.~Buchdahl, Mon.\ Not.\ Roy.\ Astron.\ Soc.\ {\bf 150}, 1 (1970).


\bibitem{08032236} V.~Acquaviva, A.~Hajian, D.~N.~Spergel, and S.~Das, Phys.\ Rev.\ D {\bf 78}, 043514 (2008) [arXiv:0803.2236].

\bibitem{08093374} R.~Gannouji, B.~Moraes, and D.~Polarski, \JCAP {\bf 0902}, 034 (2010) [arXiv:0809.3374].

\bibitem{09035296} J.~B.~Dent, S.~Dutta, L.~Perivolaropoulos, Phys.\ Rev.\ D {\bf 80}, 023514 (2009) [arXiv:0903.5296].

\bibitem{09082669} S.~Tsujikawa, R.~Gannouji, B.~Moraes, and D.~Polarski, Phys.\ Rev.\ D {\bf 80}, 084044 (2009) [arXiv:0908.2669].


\bibitem{Bergmann} P.~G.~Bergmann, Int.\ J.\ Theor.\ Phys\ {\bf 1}, 25 (1968).

\bibitem{Nordtvedt} K.~Nordtvedt, Astrophys.\ J\ {\bf 161}, 1059 (1970).

\bibitem{Wagoner} R.~Wagoner, Phys.\ Rev.\ D {\bf 1}, 3209 (1970).


\bibitem{0001066} B.~Boisseau, G.~Esposito-Farese, D.~Polarski, and A.~A.~Starobinsky, Phys.\ Rev.\ Lett. {\bf 85}, 2236 (2000) [arXiv:gr-qc/0001066]. 

\bibitem{07101510} D.~Polarski and R.~Gannouji, Phys.\ Lett.\ B {\bf 660}, 439 (2008) [arXiv:0710.1510]. 

\bibitem{08024196} R.~Gannouji and D.~Polarski, J.\ Cosmol.\ Astropart.\ Phys {\bf 0805}, 018 (2008) [arXiv:0802.4196].


\bibitem{Fradkin} E.~S.~Fradkin and A.~A.~Tseytlin, Nucl.\ Phys.\ B {\bf 261}, 1 (1985).

\bibitem{Callan} C.~G.~Callan, D.~Friedan, E.~J.~Martinec, and M.~J.~Perry, Nucl.\ Phys.\ B {\bf 262}, 593 (1985).

\bibitem{Lovelace} C.~Lovelace, Nucl.\ Phys.\ B {\bf 273}, 413 (1985).

\bibitem{Green} B.~Green, J.~M.~Schwarz, and E.~Witten, {\it Superstring Theory}, Cambridge University Press (1987).

\bibitem{Polchinski} J.~Polchinski, {\it String Theory}, Cambridge University Press (1998).


\bibitem{Teyssandier} P.~Teyssandier and P.~Tourrenc, J.\ Math.\ Phys.\ {\bf 24}, 2793 (1983).

\bibitem{Magnano} G.~Magnano, M.~Ferraris, and M.~Francaviglia, Gen.\ Rel.\ Grav.\ {\bf 19}, 465 (1987).

\bibitem{9307034} D.~Wands, Class.\ Quant.\ Grav.\ {\bf 11}, 269 (1994) [arXiv:gr-qc/9307034].

\bibitem{0307338} T.~Chiba, Phys.\ Lett.\ B {\bf 575}, 1 (2003)
[arXiv:astro-ph/0307338]. 

\bibitem{09100434} M.~Capone and M.~L.~Ruggiero, Class.\ Quant.\ Grav.\ {\bf 27}, 125006 (2010) [arXiv:0910.0434].


\bibitem{0308111} E.~E.~Flanagan, Phys.\ Rev.\ Lett.\ {\bf 92}, 071101 (2004) [arXiv:astro-ph/0308111].

\bibitem{0604028} T.~Sotiriou, Class.\ Quant.\ Grav.\ {\bf 23}, 5117 (2006) [arXiv:gr-qc/0604028].


\bibitem{0511693} J.~Larena, J.-M.~Alimi, and A.~Serna, Astrophys.\ J.\ {\bf 658}, 1 (2007) [arXiv:astro-ph/0511693].

\bibitem{0601299} A.~Coc, K.~A.~Olive, J.-P.~Uzan, and E.~Vangioni, Phys.\ Rev.\ D {\bf 73}, 083525 (2006) [arXiv:astro-ph/0601299].

\bibitem{08111845} A.~Coc, K.~A.~Olive, J.-P.~Uzan, and E.~Vangioni, Phys.\ Rev.\ D {\bf 79}, 103512 (2009) [arXiv:0811.1845].


\bibitem{0403654} V.~Acquaviva, C.~Baccigalupi, and F.~Perrotta, Phys.\ Rev.\ D {\bf 70}, 023515 (2004) [arXiv:astro-ph/0403654].

\bibitem{0412120} C.~Schimd, J.-P.~Uzan, and A.~Riazuelo, Phys.\ Rev.\ D {\bf 71}, 083512 (2005) [arXiv:astro-ph/0412120].



\bibitem{9906066} F.~Perrotta, C.~Baccigalupi, and S.~Matarrese, Phys.\ Rev.\ D {\bf 61}, 023507 (2000) [arXiv:astro-ph/9906066].

\bibitem{08032238} T.~Giannantonio, Y.-S.~Song, and K.~Koyama, Phys.\ Rev.\ D {\bf 78}, 044017 (2008)
[arXiv:0803.2238]. 

\bibitem{SLAIP} S.~Lee, AIP\ Conf.\ Proc.\ {\bf 1059}, 27 (2008). 

\bibitem{09092045} T.~Giannantonio, M.~Martinelli, A.~Silvestri, and A.~Melchiorri, \JCAP {\bf 1004}, 030 (2010)
[arXiv:0909.2045]. 

\bibitem{13081142} T.~Chiba and M.~Yamaguchi, \JCAP {\bf 1310}, 040 (2013) [arXiv:1308.1142]. 


\bibitem{0504582} L.~Perivolaropoulos, \JCAP {\bf 0510}, 001 (2005) [arXiv:astro-ph/0504582].

\bibitem{0606287} R.~Gannouji, D.~Polarski, A.~Ranquet, A.~A.~Starobinsky, \JCAP {\bf 0609}, 016 (2006) [arXiv:astro-ph/0606287].

\bibitem{0610092} S.~Nesseris and L.~Perivolaropoulos, \JCAP {\bf 0701}, 018 (2007) [arXiv:astro-ph/0610092].


\bibitem{0011115} G.~Esposito-Farese, Proceedings World Scientific 1749 (2002) [arXiv:gr-qc/0011115].

\bibitem{0107386} A.~Riazuelo and J.-P.~Uzan, Phys.\ Rev.\ D {\bf 66}, 023525 (2002) [arXiv:astro-ph/0107386].

\bibitem{0508542} S.~Tsujikawa, Phys.\ Rev.\ D {\bf 72}, 083512 (2005) [arXiv:astro-ph/0508542].

\bibitem{0612569} T.~Faulkner, M.~Tegmark, E.~F.~Bunn, and Y.~Mao, Phys.\ Rev.\ D {\bf 76}, 063505 (2007) [arXiv:astro-ph/0612569].

\bibitem{07053586} S.~Capozziello, S.~Nesseris, and L.~Perivolaropoulos, \JCAP {\bf 0712}, 009 (2007)     [arXiv:0705.3586]. 

\bibitem{08031106} S.~Tsujikawa, K.~Uddin, S.~Mizuno, R.~Tavakol, and J.~Yokoyama, Phys.\ Rev.\ D {\bf 77}, 103009 (2008) [arXiv:0803.1106]. 

\bibitem{10031686} L.~Jarv, P.~Kuusk, and M.~Saal, Phys.\ Rev.\ D {\bf 81}, 104007 (2010)
[arXiv:1003.1686]. 

\bibitem{10061246} L.~Jarv, P.~Kuusk, and M.~Saal, Phys.\ Lett.\ B {\bf 694}, 1 (2010)
[arXiv:1006.1246]. 

\bibitem{10112915} B.~Boisseau, [arXiv:1011.2915].


\bibitem{Magnano} G.~Magnano and L.~M.~Sokolowski, \PRD {\bf 50}, 5039 (1994) [arXiv:gr-qc/9312008].

\bibitem{Dick} R.~Dick, Gen.\ Rel.\ Grav. {\bf 30}, 435 (1998).

 
\bibitem{9910176} V.~Faraoni and E.~Gunzig,  Int. J. Theor. Phys. 38, 217 (1999) [arXiv:astro-ph/9910176].
 
\bibitem{0205187}  C.~Armendariz-Picon, \PRD {\bf 66}, 064008 (2002) [astro-ph/0205187].

\bibitem{0604492} R.~Catena, M.~Pietroni and L.~Scarabello, \PRD {\bf 76}, 084039 (2007) [astro-ph/0604492].

\bibitem{0605109} A.~Bhadra, K.~Sarkar, D.~P.`Datta and K.~K.~Nandi,  Mod.\ Phys.\ Lett.\ A {\bf 22}, 367 (2007) [arXiv:gr-qc/0605109].
 
\bibitem{10035394} S.~Capozziello, P.~Martin-Moruno, and C.~Rubano,  \PLB {\bf 689}, 117 (2010) [arXiv:1003.5394].
 

\bibitem{160106152} N.~Banerjee and B.~Majumder, \PLB {\bf 754}, 129 (2016) [arXiv:1601.06152].

\bibitem{0009034} G.~Esposito-Far${\rm \grave{e}}$se and D.~Polarski, \PRD {\bf 63}, 063504 (2001) [arXiv:gr-qc/0009034]. 







\end{thebibliography}
\end{document}